\documentclass[10pt,conference]{IEEEtran}
\usepackage{notation}
\usepackage{epsfig}
\usepackage{color}
\usepackage{amsmath}
\usepackage{amssymb}
\usepackage{enumerate}
\usepackage{multirow}

\begin{document}

% paper title
\title{Tree-structure Expectation Propagation for Decoding LDPC codes over Binary Erasure Channels}

% author names and affiliations
% use a multiple column layout for up to three different
% affiliations
\author{
\authorblockN{Pablo M. Olmos, Juan Jos\' e Murillo-Fuentes}
\authorblockA{Departamento de Teor\'ia de la Se\~nal y Comunicaciones.\\
Universidad de Sevilla.\\
email:\tt \{olmos,murillo\}@us.es}
\and
%\authorblockN{Juan Jos\' e Murillo-Fuentes}
%\authorblockA{Departamento de Teor\'ia de la Se\~nal \\ y Comunicaciones\\
%Universidad de Sevilla.\\
%Email: murillo@us.es}
%\and
\authorblockN{Fernando P\'erez-Cruz
}
\authorblockA{Departamento de Teor\'ia de la Se\~nal y Comunicaciones.\\
Universidad Carlos III Madrid.\\
email: \tt fernando@tsc.uc3m.es
}

}

% make the title area
\maketitle

\begin{abstract}
Expectation Propagation is a generalization to Belief Propagation (BP) in two ways. First, it can be used with any 
exponential family distribution over the cliques in the graph. Second, it can impose additional constraints on the 
marginal distributions. We use this second property to impose pair-wise marginal distribution constraints in some 
check nodes of the LDPC Tanner graph. These additional constraints allow decoding the received codeword when the BP 
decoder gets stuck. In this paper, we first present the new decoding algorithm, whose complexity is identical to the 
BP decoder, and we then prove that it is able to decode codewords with a larger fraction of erasures, as the block size 
tends to infinity. The proposed algorithm can be also understood as a simplification of the Maxwell decoder, but without 
its computational complexity. We also illustrate that the new algorithm outperforms the BP decoder for finite block-size codes.
%The Maxwell decoder has been proposed for bridging the gap between the achievable capacity by belief propagation decoding
%and the maximum a posteriori decoding in binary erasure channels for LDPC codes. The Maxwell decoder, once
%the belief-propagation decoder gets stuck in a nonempty stopping set, guesses a bit and replicates any running copy of the
%decoding process. Density evolution and EXIT chart analyses of this iterative decoder show that MAP performance can be explained
%from the performance of the BP decoder. The complexity of the Maxwell decoder depends exponentially on the number of guesses and
%a priori we cannot bound the number of guesses, which limits its applicability as an LDPC decoder. In this paper, we adapt the
%expectation propagation algorithm for LDPC decoding in the BEC. Our algorithm can be understood as a Maxwell decoder with a
%complexity similar to belief propagation. We include a detailed analysis to demonstrate that the
%achieved capacity is higher than that of the belief propagation decoder.
\end{abstract}

\section{Introduction}

In this paper we propose a new algorithm to decode low-density parity-check (LDPC) codes over the binary erasure
channel (BEC). The proposed algorithm has a better performance than the traditional Belief Propagation (BP) decoder with the
same computational complexity. The analysis of the BP decoder over independent and identically distributed BEC with erasure
probability $\epsilon$ has been detailed in \cite{Luby01,Urbanke02,Oswald02}, in which its limiting performance and optimization
are addressed. Once we represent the LDPC code by the bipartite graph induced by
its parity-check matrix (Tanner Graph \cite{Tanner81}), the BP decoding algorithm can be easily described. The degree of a check
 node is the number of variable nodes connected to it in the bipartite graph, and vice-versa.

We initialize the decoder by removing from the graph all the variable nodes corresponding to non-erased bits. We also remove all the connections from these variable nodes.
After removing a variable node whose value was one, we change the parity of the check node(s) it was connected to. After the initialization stage,
the BP algorithm proceeds by removing a check node and a variable node in each step:
\begin{enumerate}
\item It looks for any check node linked to a single variable node (a check node of degree one). The BP decoder copies the
 parity of this check node into the variable node and removes the check node.
\item Second, it removes the variable node that we have just de-erased. If the variable was a one,
it changes the parity of the check node(s) it was connected to.
\item It repeats Steps 1 and 2 until all the variable nodes have been removed, successfully finishing the decoding of the
received word, or until there are no degree-one check nodes left, yielding an unsuccessful decoding.
\end{enumerate}

The asymptotic analysis of the BP performance decoding LDPC codes over the BEC is derived in \cite{Luby01,Urbanke01-2} and 
capacity-achieving degree distributions are presented in \cite{Luby01,Oswald02}. However, for graphs with cycles, the performance of the BP decoder 
for any finite-length code is always inferior to the maximum a posteriori (MAP) decoder. For example, in \cite{Urbanke08} 
it is shown that a BP decoder for a rate-$1/2$ regular LDPC code with three ones per column and six ones per row can
decode channels with erasure probability up to $\pe=0.4294$, which is typically denoted as the BP capacity, as the block size tends to infinity, while the limiting MAP
 performance reaches $0.48815$, which is denoted as the MAP capacity.

In \cite{Urbanke08}, the authors proposed the Maxwell decoder to achieve the MAP solution for decoding LDPC codes over BEC. Once
 the BP gets stuck, because there are no additional degree-one check nodes, the Maxwell decoder assumes that the bits of a set 
 of variable nodes are known so there is one degree-one check node that allows restarting the BP algorithm. This step is repeated as many times as needed, until all variable nodes have been accounted for. The complexity of the Maxwell decoder grows exponentially with the number of guessed variables and there is no meaningful upper bound on how many we need to assume known. Hence, it is an impractical algorithm to decode LDPC codes over BEC, although it is a useful tool to examine the code performance and derive the MAP capacity \cite{Urbanke08}. The idea of a BP decoder with variable guessing was first proposed in \cite{Pishro04} for short-length LDPC codes, where the complexity of the algorithm is relatively small.

In this paper, we propose a new algorithm that is able to continue decoding LDPC codes for BEC after an unsuccessful
BP decoding. This algorithm improves the performance of the BP decoder for finite length codes at a similar complexity.
We include the analysis of the asymptotic case to show that the achieved capacity is higher than the BP one. Our decoding algorithm borrows from the Tree-structured approximations for Expectation 
Propagation \cite{Minka03} and we refer to it as the TEP algorithm. The algorithm in \cite{Minka03} can be understood as a generalization 
of the BP decoder, in the sense that the variable nodes are linked to impose some pair-wise marginal constraints, instead 
of trying to compute only independent marginal distributions.

The rest of the paper is organized as follows. Section\SEC{TEP} is devoted to introducing the \TEP algorithm for decoding LDPC codes. In Section\SEC{Analysis}, we analyze a simplified version of the \TEP decoder to prove that it achieves a higher capacity than BP. In Section\SEC{Exp}, we compare the performance of the \TEP and BP decoders with finite-length regular LDPC codes. We conclude in Section VI with some final comments.

\section{\TEP algorithm}\LABSEC{TEP}

The TEP decoder starts once the BP gets stuck. In a way, it works as the Maxwell decoder in which some variable nodes are assumed to be known to continue decoding. But the \TEP complexity is identical to the BP, as it removes a check node and a variable node in each step and its complexity does not depend on the number of assumed variable nodes.

The Tanner graph of the LDPC code, denoted as $\codegraph$, has $\n$ variable nodes $\Vjoin{1},\ldots,\Vjoin{\n}$
and $\n(1-\rate)$ check nodes $\CN{1},\ldots,\CN{\n(1-\rate)}$, where $r$ is the rate of the code. The degree of a variable
node $\Vjoin{m}$ and a check node $\CN{j}$ is, respectively, denoted by $\vd{m}$ and $\cd{j}$. The graph $\codegraph$ is
reduced by removing the non-erasured variables and performing the BP decoder until there are no degree-one check nodes left. We
refer to the graph at the end of the BP as $\RBP$. The \TEP decoder works over $\RBP$ using the degree-two check nodes. A check
 node of degree two tells us that the variable nodes connected to it are either equal, if the check has parity zero, or different 
 otherwise. Our algorithm chooses any check node with degree two and removes it from the graph together with one of the variable nodes connected to it and the two associated edges.
Then it reconnects to the remaining variable node all the check nodes that were connected to the removed variable node. Finally, the
 parities of the check nodes connected to the remaining variable node have to be reversed if the removed degree-two
check node had parity one. We sketch the procedure in Fig. \FIG{Fig1}, where the algorithm selects the check node $\CN{1}$
and removes it along with the variable $\Vjoin{2}$. We can observe in Fig. \FIG{Fig1}(b) that $\Vjoin{1}$ is now connected to the check nodes $\CN{2}$ and $\CN{3}$
and its degree is now $\vd{1}+\vd{2}-2$.

% If the parity of $\CN{1}$ is one, we have to reverse the parity of the check
%nodes $\CN{2}$ and $\CN{3}$.

\begin{figure}
\centering
\begin{tabular}{cc}
\includegraphics[width=4 cm]{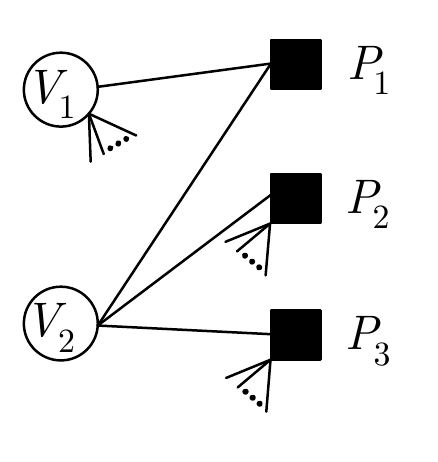} & \includegraphics[width=4 cm]{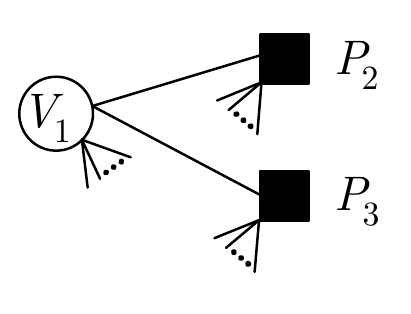}\\
(a) & (b)
\end{tabular}
\caption{In (a) we show two variable nodes, $\Vjoin{1}$ and $\Vjoin{2}$, that share a check node of degree two. In (b), we show the graph once $\CN{1}$ and $\Vjoin{2}$ have been removed. If $\CN{1}$ is parity one, the parities of $\CN{2}$ and $\CN{3}$ are reversed.}\LABFIG{Fig1}
\end{figure}

%We just have to list the variables that are equal or different to the remaining variables in the graph.
%The decoder only has to store a list of the removed variables and the associated remaining variables in the graph.

The algorithm removes a check and a variable node per iteration, as the BP does. The removal of a check node and a variable does
not increase the complexity of the decoder, as it happens in the Maxwell decoder \cite{Urbanke08}.
Unlike the BP, the value of the removed variable node is unknown. This variable becomes known, when the remaining variable node
in the graph has been decoded. Hence, the decoder only has to store a list of the removed variables and the associated remaining
variables in the graph. In the example of Fig. \FIG{Fig1}, it means that $V_2$ would be known once $V_1$ has been de-erased. 

This procedure continues removing check and variable nodes, but it does not set any of the
variables. To be able to set any variable node, we need to find a check node with degree two such that the two variable nodes
connected to it also share another check node with degree three, as illustrated in Fig. \FIG{Fig2}(a). When we remove the check
node $\CN{3}$ and the variable node $\Vjoin{2}$, the check node $\CN{4}$ is now degree one, as illustrated in Fig. \FIG{Fig2}(b),
and we can restart the BP algorithm.
%
%In this case, $\Vjoin{1}$ and $\Vjoin{2}$ share two check nodes in Fig. \FIG{Fig2}(a) and the degree of $\Vjoin{2}$ in Fig. \FIG{Fig2}(b) is $\vd{1}+\vd{2}-4$. In the example illustrated in Fig. \FIG{Fig1}(a), the variables connected to the check node of degree $2$ do not share another check node. Hence, no check node of degree $1$ can be created and the degree of the remaining variable in Fig. \FIG{Fig1}(b) is $\vd{1}+\vd{2}-2$.

%\begin{figure}[h]
%\centering
%\begin{tabular}{cc}
%\includegraphics[width=4 cm]{figuresv2/NewFig2.pdf} & \includegraphics[width=4 cm]{figuresv2/NewFig2b.pdf}\\
%(a) & (b)
%\end{tabular}\caption{In (a), the variables $\Vjoin{1}$ and $\Vjoin{2}$ are connected to a two-degree check node, $\CN{3}$, and they also share a check node of degree three, $\CN{4}$.
%In (b) we show the graph once the algorithm has removed $\CN{3}$ and $\Vjoin{2}$.}\LABFIG{Fig2}
%\end{figure}

\begin{figure}[h]
\centering
\begin{tabular}{cc}
\includegraphics[width=4 cm]{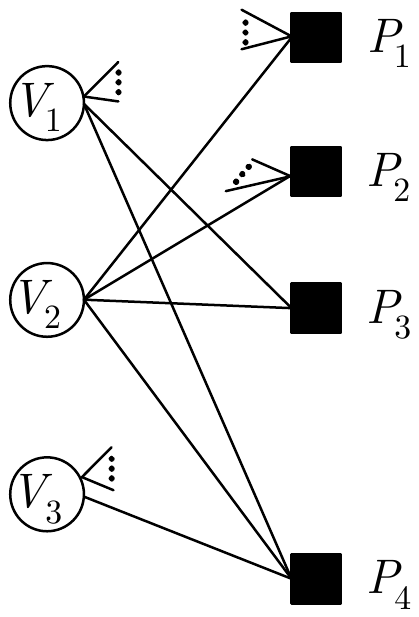} & \includegraphics[width=4 cm]{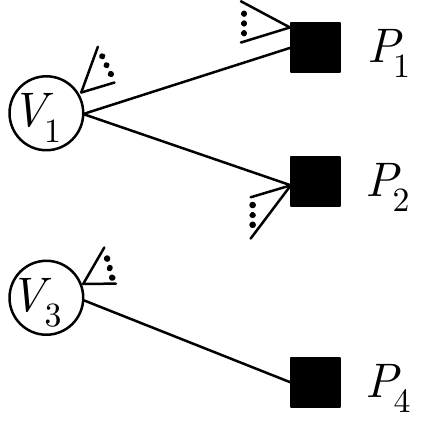}\\
(a) & (b)
\end{tabular}\caption{In (a), the variables $\Vjoin{1}$ and $\Vjoin{2}$ are connected to a two-degree check node, $\CN{3}$, and they also share a check node of degree three, $\CN{4}$.
In (b) we show the graph once the algorithm has removed $\CN{3}$ and $\Vjoin{2}$.}\LABFIG{Fig2}
\end{figure}

When the \TEP algorithm starts running over the graph $\RBP$, it is very unlikely that the two variable nodes in a check node of degree two also share a check node of degree three. However, as we remove variable and check nodes the probability of this happening grows, since we are reducing the number of nodes in the graph and increasing the degree of the remaining variable nodes. In the following section, we prove that the growth of this probability is exponential with the number of iterations of the TEP decoder. In each iteration we remove a check node of degree two and a variable node connected to it. Eventually, when a check node of degree one is created, we can restart the BP algorithm and correct error patterns undecodable by the BP decoder. The \TEP decoder halts either when the graph runs out of check nodes of degree one and two or when all the variables have been de-erased.

%In the first case, the \TEP final graph is denoted as $\RTEP$.

\section{Analysis of the \TEP decoder}\LABSEC{Analysis}

We now analyze a simplification of the above procedure and show that it achieves a higher capacity than BP. The \TEP decoder first removes all the check nodes of degree two, as explained in Section\SEC{TEP}, and then it runs the BP decoder. This modification allows applying the results
in \cite{Luby01}, in which the BP capacity for BEC was computed, to analyze the TEP decoder. We consider a LDPC ensemble $\LDPC$ defined by the code length $\n$ and the edge degree distribution (DD) pair $[\lamb{x},\rh{x}]$ \cite{Luby01-2}:
\begin{eqnarray}\LABEQ{ensemble}
&&\lambfull,\\
&&\rhofull,
\end{eqnarray}
where $\lambi{i}$ represents the fraction of edges with left degree $i$ in the graph and $\rhi{i}$ is the fraction of edges with right degree $i$. The left
(right) degree of an edge is the degree of the variable (check) node it is connected to. The total number of edges in the graph
is denoted by $\Edges$ and it can be readily check that
\begin{equation}
\Edges=\frac{\n}{\sum_{i}\lambi{i}/i}.
\end{equation}
%Remark that the graph is specified in terms of fractions of edges, and not nodes, of each degree; this form is more convenient to derive the capacity of the BP decoder, but also to analyze the \TEP decoder as we show later.
The passage of time during the decoding process is scaled so that each time unit of length $\Inctime$ corresponds to one
 iteration of the decoding process as in \cite{Luby01}. In one iteration, the \TEP removes a check node of degree two and one variable.

Let define  $\li{i}{t}$ and $\ri{i}{t}$ such that $\Li{i}{t}=\li{i}{t}\Edges$ and $\Ri{i}{t}=\li{i}{t}\Edges$ are, respectively, the number of edges with left and right degree $i$ in the graph at time $t$. We denote $\eremain{t}$ as the number of edges in the graph at time $t$ divided by $\Edges$. Hence, $\li{i}{t}/\eremain{t}$ for $i=1,\ldots,\lambimax$ and $\ri{i}{t}/\eremain{t}$ for $i=1,\ldots,\rhimax$ are the coefficients of the DD pair that
defines the graph at time $t$. In the analysis of the BP decoder in \cite{Luby01}, the evolution of the \dd pair in the limiting case, as $\n$ tends to infinity, is described by a set of differential equations. The BP decoder has analytical solution, which holds while $\ri{1}{t}>0$. In particular, the solution for $\ri{1}{t}$ is given by:
\begin{equation}\LABEQ{r1solution}
\ri{1}{x}=\pe\lamb{x}\left[x-1+\rh{1-\pe\lamb{x}}\right].
\end{equation}
The BP capacity, $\peBP$, is the maximum probability of erasure such that $\ri{1}{x}>0\;\forall x\in(0,1]$. 
%We show that, when the \TEP starts,
%the probability that the two variable nodes in a check node of degree two also share another check node, as illustrated in
%Fig. \FIG{Fig2}(a), is zero. This allows us to split the analysis into two stages.

\subsection{$\TEP$ analysis}

For a BEC with $\pe>\peBP$, the solution of $\ri{1}{x}$ in \EQ{r1solution} touches $0$ for $x=\xBP>0$  and the BP decoder gets stuck at this point. The residual graph is $\RBP$ and it is defined by the \dd pair
\begin{equation}\LABEQ{BPdd}
\left[\sum_{i}\frac{\li{i}{\xBP}}{\eremain{\xBP}}x^{i-1}, \sum_{i}\frac{\ri{i}{\xBP}}{\eremain{\xBP}}x^{i-1},\nBP\right],
\end{equation}
where $\nBP$ is the number of remaining variables.  The \TEP decoder takes the graph $\RBP$ as input and removes one check node of degree two and one variable per iteration. While in the BP decoder, the basic step has an unique formulation, now we have two different situations. Once a check node $\CN{j}$ of degree $\cd{j}=2$ has been found:
\begin{enumerate}[A.]
\item The variable nodes connected with the check, $\Vjoinone$ and $\Vjointwo$, do not share another check node. See Fig. \FIG{Cases}(a).
\item $\Vjoinone$ and $\Vjointwo$ share another check node $\CN{l}$ of degree $\cd{l}=m$, where $m\in\{2,\ldots,\rhimax\}$. See Fig. \FIG{Cases}(b).
\end{enumerate}

%\begin{figure}
%\centering
%\begin{tabular}{cc}
%  \includegraphics[width=4 cm]{figuresv2/NewFig3a.pdf} & \includegraphics[width=4 cm]{figuresv2/NewFig3b.pdf} \\
%  (a) & (b)
%\end{tabular}
%\caption{Given the check node $\CN{j}$, in (a) the variable nodes $\Vjoinone$ and $\Vjointwo$ share only the check node $\CN{j}$. In (b), they share another one of degree $m\in\{1,\ldots,\rhimax\}$.}\LABFIG{Cases}
%\end{figure}

\begin{figure}
\centering
\begin{tabular}{cc}
  \includegraphics[width=4 cm]{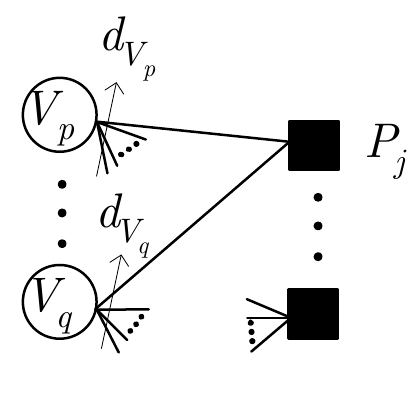} & \includegraphics[width=4 cm]{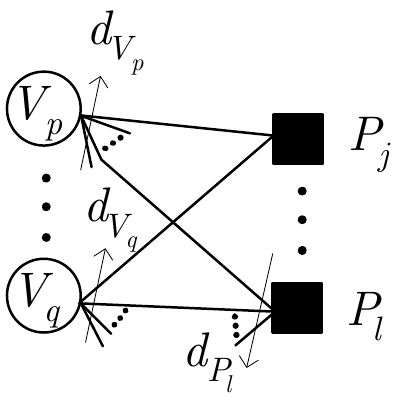} \\
  (a) & (b)
\end{tabular}
\caption{Given the check node $\CN{j}$, in (a) the variable nodes $\Vjoinone$ and $\Vjointwo$ share only the check node $\CN{j}$. In (b), they share another one of degree $m\in\{1,\ldots,\rhimax\}$.}\LABFIG{Cases}
\end{figure}

In the ensemble in \EQ{BPdd} and provided that $\Vjoinone$ and $\Vjointwo$ are connected to $\CN{j}$ of degree two, $\pB{\xBP}$ represents the probability that the variable nodes also share another check node. This probability can be approximated as follows:
\begin{eqnarray}\LABEQ{pB}
\pB{\xBP}\approx\sum_{m=2}^{\rhimax}\frac{(\lav{\xBP})^{2}\ri{m}{\xBP}}{\eremain{\xBP}\Edges},
\end{eqnarray}
where $\lav{\xBP}=\sum_{i}i\li{i}{\xBP}/\eremain{\xBP}$ is the average edge left degree for the ensemble in \EQ{BPdd}.
This expression is proven in \cite{Olmos10}. In the limiting case, as $\Edges\rightarrow\infty$, then $\pB{\xBP}\rightarrow0$.
%Hence, it is very unlikely that check nodes of degree one can be created at the beginning of the \TEP decoding.

We have divided the \dd evolution during the decoding process in two stages, each one described by a set of differential
equations:
\begin{itemize}
\item Stage A. The equations describe the \dd evolution assuming that, in one iteration of the \TEP decoder, the
variable nodes connected to any check node of degree two do not share another check node.
\item Stage B. The equations describe the \dd evolution when, in one iteration of the \TEP decoder, the variable
nodes connected to a check node of degree two always share another check node.
\end{itemize}

The Stage A equations model the evolution of the \dd pair at the beginning of the \TEP decoding. The solution of the equations
shows the probability $\pB{t}$ grows exponentially with time. Although the Stage A equations assume $\pB{t}$ zero or negligible,
we use them to approximate the evolution of the \dd pair until $\pB{t}=1$. Notice that this is a conservative solution in the
sense we are ignoring the check nodes of degree one that could be created before $\pB{t}$ is one. The Stage A lasts between $t=0$,
when the \TEP starts, and $t=\tA$. We define $\tA$ as the time in which $\pB{t}$ is either $1$ or the graph runs out of check nodes of degree two. In the first case, 
Stage B starts. In the second one, the \TEP decoder finishes without creating check nodes of degree one and
it cannot improve the BP solution. 

In the Stage B, $\pB{t}=1$ and the variables connected to a check node of degree two also share another check node. Hence, check 
nodes of degree one can be created. The Stage B equations hold until the decoder removes all the check nodes of degree two. At this 
time, denoted as $\tB$, the BP decoder is run again. 

%Remark than in Stage B, the graph verifies $\pB{t}=1$ for $t<\tB$, which means that
%two variables connected to a check node of degree two also share another check node and, hence, check nodes of degree one can
%be created.

%At $\tA$, if there are check nodes of degree two left, Stage B starts. As $\pB{t}=1$, two variables connected to a check node of
%degree two also share another check node and, hence, check nodes of degree one can be created. The Stage B finishes when the
%graph has no check nodes of degree two. At this time, denoted as $\tB$, the BP decoder is run again.

%When Stage A finishes, if there are check nodes of degree two left, Stage B starts. The \TEP decoder ends when it
% removes all the check nodes of degree two.
%The Stage B starts at $t=\tA$ if $\pB{t}=1$ and $\ri{2}{t}>0$. Let $\tB$ represents the time when the graph has no check nodes
%of degree two. At this time, the BP decoder is run again. Remark that it is in the Stage B when check nodes of degree one are
%created.

\subsection{Stage A of decoding} \LABSEC{StageA}

%In this subsection, we derive the expected variation of the  $\Li{i}{t}$ and $\Ri{i}{t}$ distributions in one iteration of \TEP decoder assuming that each time it finds a check node of degree two, the variable nodes connected to it do not share another check node.

Suppose that, at time $t$, we remove the check node $\CN{j}$ in Fig. \FIG{Cases}(a), which is connected to
$\Vjoinone$ and $\Vjointwo$. If $\Vjoinone$ is the remaining variable, its degree is now $\vd{\indone}+\vd{\indtwo}-2$. From the
 edge perspective, the graph losses $\vd{\indone}$ edges with left degree $\vd{\indone}$ and $\vd{\indtwo}$ edges with left degree
 $\vd{\indtwo}$ and gains $\vd{\indone}+\vd{\indtwo}-2$ edges with left degree $\vd{\indone}+\vd{\indtwo}-2$. The probability
 that $\vd{\indone}$ or $\vd{\indtwo}$ are $i$ is $\li{i}{t}/\eremain{t}$ for $i=1,\ldots,\Tlmax{t}$. 
 
% Since we are increasing
% the degree of variable nodes, we have to include a dependence on time in the maximum left degree, $\Tlmax{t}$.

To derive the differential equation that describes the evolution of $\li{i}{t}$ for $i=1,\ldots,\Tlmax{t}$, we compute the expected variation
of the $\Li{i}{t}$ distribution, $\E\left[\Li{i}{t+\Inctime}-\Li{i}{t}\right]$. Hence, we are interested in cases where
we have $\Li{i}{t+\Inctime}\neq\Li{i}{t}$. For $i\geq3$:

%When $\CN{j}$ and $\Vjoinone$ are removed, the distribution $\Li{i}{t+\Inctime}$ for $i=2,\ldots,\Tlmax{t+\Inctime}$ is modified with respect to $\Li{i}{t}$ depending on $\vd{\indone}$ and $\vd{\indtwo}$. For $i\geq3$, in three cases we have $\Li{i}{t+\Inctime}\neq\Li{i}{t}$:
\begin{enumerate}
\item $\Li{i}{t+\Inctime}=\Li{i}{t}-i$ if $(\vd{\indone}=i$ XOR  $\vd{\indtwo}=i)$
AND $\vd{\indone}+\vd{\indtwo}-2\neq i$.
\item $\Li{i}{t+\Inctime}=\Li{i}{t}-2i$ if $ \vd{\indone}=\vd{\indtwo}=i$.
\item $\Li{i}{t+\Inctime}=\Li{i}{t}+i$ if $\vd{\indone}\neq i, \vd{\indtwo}\neq i$ AND $ \vd{\indone}+\vd{\indtwo}-2=i$.
\end{enumerate}

The probability of each case is respectively,
\begin{eqnarray}
&p_{1,i}&=2\frac{\li{i}{t}}{\eremain{t}}\left(1-\frac{\li{i}{t}}{\eremain{t}}-\frac{\li{2}{t}}{\eremain{t}}\right),\LABEQ{prob1A}\\&p_{2,i}&=\left(\frac{\li{i}{t}}{\eremain{t}}\right)^{2},\LABEQ{prob2A}\\&p_{3,i}&=\convfull{2}{i}\LABEQ{prob3A}.
\end{eqnarray}
For $i=2$, $\Li{2}{t+\Inctime}\neq\Li{2}{t}$ only in Cases $1$ and $3$. The expected variation of the $\Li{i}{t+\Inctime}$ distribution with respect to $\Li{i}{t}$ is
\begin{eqnarray}
&&\E\left[\Li{2}{t+\Inctime}-\Li{2}{t}\right]=-2(p_{1,2}-p_{3,2})\LABEQ{ExpLiA1},\\
&&\E\left[\Li{i}{t+\Inctime}-\Li{i}{t}\right]=-i(p_{1,i}+2p_{2,i}-p_{3,i}).\;\;\;\; \LABEQ{ExpLiA2}
\end{eqnarray}

If we set $\Inctime=\Edges^{-1}$ and $\n\rightarrow\infty$, then \EQ{ExpLiA1} and \EQ{ExpLiA2} can be rewritten as \cite{Luby01}:
\begin{eqnarray}
&&\frac{\partial\li{2}{t}}{\partial t}=-2\left(p_{1,2}-p_{3,2}\right), \qquad i=2,\LABEQ{difliA1}\\
&&\frac{\partial\li{i}{t}}{\partial t}=-i\left(p_{1,i}+2p_{2,i}-p_{3,i}\right),\qquad i>3\LABEQ{difliA2}.
\end{eqnarray}
The initial conditions for $t=0$ are given by the BP ensemble defined in \EQ{BPdd}. The maximum
left degree $\Tlmax{t}$ grows according to $\Tlmax{t+\Inctime}=2\Tlmax{t}-2$ and, hence, $\Tlmax{t}$ grows
 as $\mathcal{O}(2^{t})$. In each iteration, two edges of right degree two are lost. Therefore, $\E\left[\Ri{2}{t+\Inctime}-\Ri{2}{t}\right]=-2$ and the differential equation for $\ri{2}{t}$ is
\begin{equation}\LABEQ{difriA}
\frac{\partial\ri{2}{t}}{\partial t}=-2\Rightarrow\ri{2}{t}=\ri{2}{0}-2t=\ri{2}{\xBP}-2t.
\end{equation}
Equations \ref{eq:difliA1} and \ref{eq:difliA2} represent a system of coupled nonlinear differential equations whose
solution has to be computed by numerical methods. The solution shows that the average left degree
$\lav{t}=\sum_{i}i\li{i}{t}/\eremain{t}$ grows exponentially with time and so does in \EQ{pB}. 

% In $\EQ{difriA}$, we can see that $\ri{2}{t}=0$ at $\tmaxA\doteq\ri{2}{\xBP}/2$. At $t=\tA$, Stage A finishes. If $\tA=\tmaxA$,
% the \TEP decoder finishes without creating any check node of
% degree one. If $\tA<\tmaxA$, the decoder enters in Stage B. This condition is related to the channel
% probability of erasure $\pe$, and it provides a maximum $\pe$ such that the decoder enters in Stage B. This maximum $\pe$ is referred as $\peA$.

%Stage A ends if the graphs runs out of check nodes of degree two and $\pB{t}\neq1$, the \TEP decoder finishes without
%creating any check node of degree one. In $\EQ{difriA}$, we can see that $\ri{2}{t}=0$ at $\tmaxA\doteq\ri{2}{\xBP}/2$. The
%Stage A also ends if  $\pB{t}$ reaches $1$ for $\tA<\tmaxA$.  This condition is related to the channel probability of erasure
%$\pe$, and it provides a maximum $\pe$ such that the decoder enters in Stage B. This value is referred as $\peA$.

Stage A can either finish, because it runs out of degree two check nodes or because all the variables nodes 
that share a check node of degree two also share another check node, which means that $\pB{t}=1$.  If the end time
 of Stage A, denoted as $\tA$, is $\tmaxA\doteq\ri{2}{\xBP}/2$, then $\ri{2}{t}$ in \EQ{difriA} is zero and we cannot create 
 new degree one-check nodes. Otherwise, if $\tA<\tmaxA$, we enter Stage B, in which the decoder creates new degree-one check nodes and 
 we can restart the BP decoder. 
 
 %The condition $\tA<\tmaxA$ is related to the channel probability of erasure
%$\pe$, and it results in a maximum $\pe$ such that the decoder enters in Stage B. This value is referred as $\peA$.

%
%
% "La Stage A termina si se agotan los check nodes de grado dos y la PB(t) no es uno. De (14) se concluye que esto ocurre para t_max,A... El Stage A tambiŽn termina si se llega PB(t)=1para un t=t_A<t_max,A. El caso li«mite en el que la PB(t) llega a uno para t_max,A viene dado por un epsilon_max,A, que dar‡ la capacidad del TEP. "

\subsection{Stage B}

The graph at the beginning of this stage is defined by the solution
of \EQ{difliA1}, \EQ{difliA2} and \EQ{difriA} at $t=\tA$. We are in a situation similar to Fig. \FIG{Cases}(b) and we select a check node as $\CN{j}$.
The variables
 $\Vjoinone$ and $\Vjointwo$ are also linked to $\CN{l}$ with degree $\cd{l}=m$. If $\vd{\indone}$ and $\vd{\indtwo}$ are the
 degree of the variables, the remaining variable node has degree $\vd{\indone}+\vd{\indtwo}-4$
 \footnote{The probability that $\Vjoinone$ and $\Vjointwo$ share a third check node $\CN{u}$ in negligible.}.
 The check node $\CN{l}$ losses two edges and its degree reduces to $\cd{l}=m-2$. The equations describing the evolution
 of the $\li{i}{t}$ distribution for $i=2,\ldots,\Tlmax{t}$ can be derived as in Section\SEC{StageA}. Regarding
  to the $\ri{m}{t}$ distribution, the graph looses $m$ edges of right degree $m$ and gains $m-2$ edges of right degree $m-2$. As
   the check node $\CN{l}$ has degree $\cd{l}=m$ with probability $\ri{m}{t}/\eremain{t}$, the $\ri{m}{t}$ evolution can
   be described as follows \cite{Olmos10}:
 \begin{eqnarray}
 &&\frac{\partial\ri{1}{t}}{\partial t}=\frac{\ri{3}{t}}{\eremain{t}},\\
 &&\frac{\partial\ri{2}{t}}{\partial t}=2\left(\frac{\ri{4}{t}}{\eremain{t}}-\frac{\ri{2}{t}}{\eremain{t}}\right)-2,\LABEQ{riB2}\\
 &&\frac{\partial\ri{m}{t}}{\partial t}=m\left(\frac{\ri{m+2}{t}}{\eremain{t}}-\frac{\ri{m}{t}}{\eremain{t}}\right)\;\;\;m>2,\LABEQ{riB1}
 \end{eqnarray}\LABEQ{riB3}
where the term $-2$ in \EQ{riB2} represents the lost edges of the check node $\CN{j}$. Let $\tB$ represents the time
when $\ri{2}{t}=0$. At $t=\tB$, all the check nodes of degree two have been removed. Stage B ends and the BP decoder
is begins. The BP input is the \dd solution of Stage B equations at $\tB$.

%This distribution is specially favorable for the BP decoder: it has a certain number of check nodes of degree one that can be connected to variable nodes of arbitrary high degree. When the BP decoder removes one of these variables, the number of edges removed is very high and also the probability that new check nodes of degree one are created.

\subsection{\TEP decoder solution for a regular LDPC ensemble}\LABSEC{RegularEx}

In this section, we estimate the \TEP decoder capacity, denoted as $\pe_{TEP}$, for a regular LDPC ensemble with $\lamb{x}=x^{2}$ and
$\rh{x}=x^{5}$. The BP capacity is $\peBP=0.4294$ \cite{Luby01,Urbanke08}. We solve \EQ{difliA1}, \EQ{difliA2} and \EQ{difriA} 
using Euler's method and compute $\epsilon_{\max,A}=0.4315$, which is the maximum channel probability of
 erasure such that the distribution has $\pB{t}=1$ before the graph runs out of check nodes of degree two.

For $\peA$ and the solution of Stage A equations at $\tA$, we solve the Stage B equations. The Stage B equations hold until the
graph has no check nodes of degree two, which happens at $t=\tB$. The \dd pair at $t=\tB$ is specially favorable for the BP decoder: it has a certain number of check nodes of degree one that
  can be connected to variable nodes of arbitrary high degree. When the BP decoder removes one check node of degree one and a
  variable, the number of edges removed is very high and also the probability that new check nodes of degree one can be created.
  As expected, the BP solution, given the input ensemble defined by the \dd at $t=\tB$, verifies the condition
  $\ri{1}{x}>0\; \forall x\in(0,1]$. Hence, the \TEP algorithm for $\n\rightarrow\infty$ can decode at least up to
  $\pe=0.4315$. This value is just a lower bound of the capacity of the \TEP decoder, since the decoder actually creates more
  check nodes of degree one than the ones provided by Stage B solution. In any case, we have shown that
  $\pe_{TEP}\geq0.4315>\peBP$.

\section{Experimental results for finite-length codes}\LABSEC{Exp}

In this section, we illustrate the performance of the \TEP decoder, as proposed in Section\SEC{TEP}, to show that
it significantly improves the performance of BP decoder for finite-length codes. We have used the regular LDPC 
ensemble analyzed in the above section.  In Fig. \FIG{WERregular} we depict the worderror rate (WER) for the \TEP decoder, solid
 lines, and the BP decoder, dashed lines, for code lengths $\n=2^{i}$ with $i=9(\circ),10(+),11(\bigtriangledown), 12(\square)$. Each curve has been averaged for 100 different samples 
 of the regular ensemble. The TEP decoder always improves the BP decoder.

%\begin{figure}
%\centering
%\includegraphics[width=9cm]{figuresv2/Tree_LDPC_CurvaWER(log).pdf}\LABFIG{WERregular}
%\caption{WER performance of the \TEP decoder (solid line) and the BP decoder (dashed line) for a regular (3,6) LDPC ensemble and code lengths $\n=2^{i}$ with $i=9(\circ),10(+),11(\bigtriangledown), 12(\square)$. Each curve has been averaged for 100 different samples of the ensemble.}
%\end{figure}

\begin{figure}
\centering
\includegraphics[width=9cm]{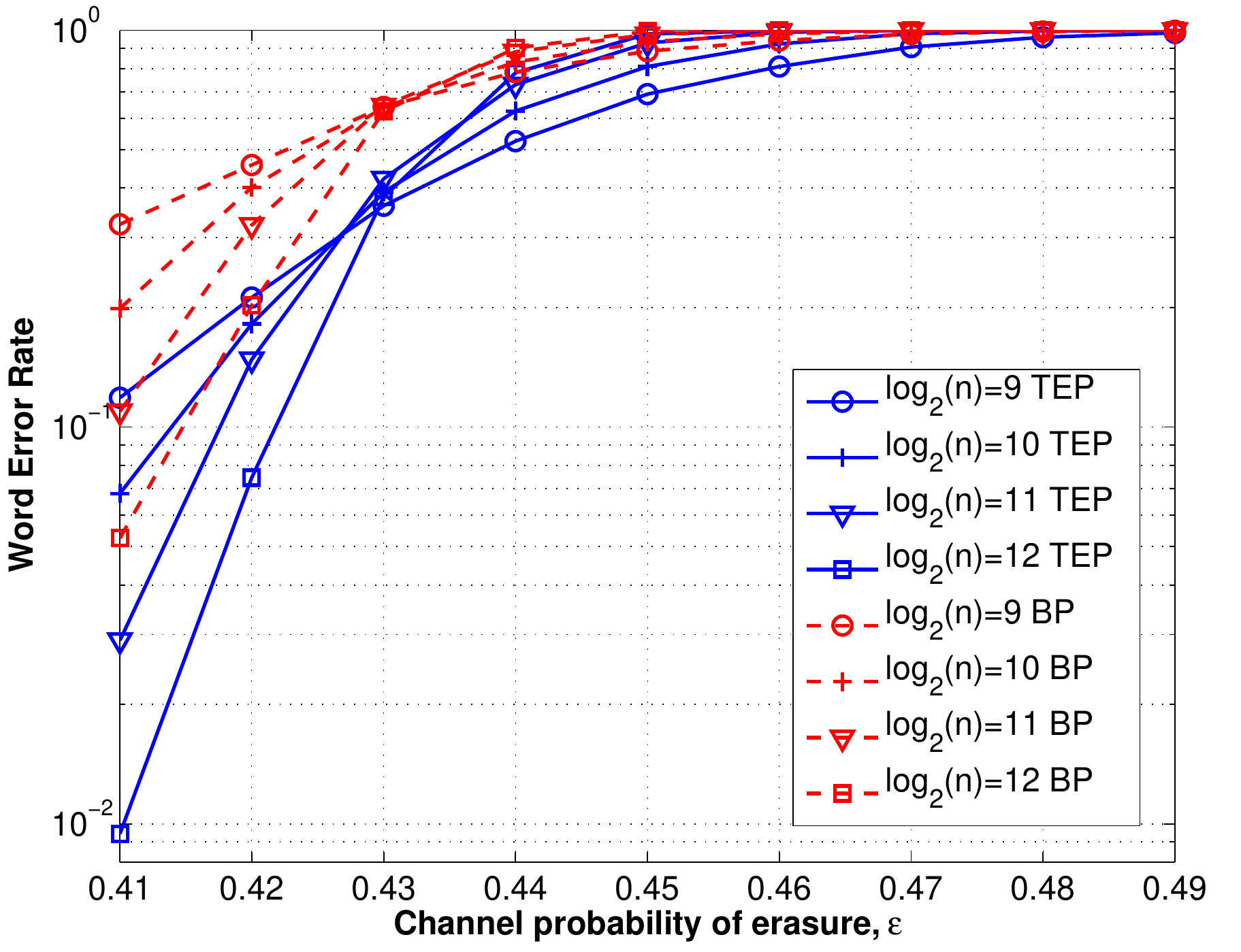}\LABFIG{WERregular}
\caption{WER performance of the \TEP decoder (solid line) and the BP decoder (dashed line) for a regular (3,6) LDPC ensemble and code lengths $\n=2^{i}$ with $i=9(\circ),10(+),11(\bigtriangledown), 12(\square)$. Each curve has been averaged for 100 different samples of the ensemble.}
\end{figure}

\section{Conclusions}

In this paper, we have proposed a new decoding algorithm for LDPC codes over the BEC.
The \TEP algorithm exhibits a significant improvement compared to the BP decoder with identical computational complexity. The \TEP algorithm borrows from
the Tree-structured approximations for Expectation Propagation and it is able to continue decoding when BP halts,, because there 
are no degree-one check nodes left. We have
proved that the \TEP decoder achieves a higher capacity than the BP by deriving the equations for the \dd
evolution along the decoding process. We have illustrated the results for regular
LDPC codes. The extension of the analysis to irregular LDPC ensembles is immediate and it is carried out in \cite{Olmos10}.

\section*{Acknowledgement}
This work was partially funded by Spanish government (Ministerio de Educaci\'on y Ciencia TEC2009-14504-C02-\{01,02\}, Consolider-Ingenio 2010 CSD2008-00010), and the European Union (FEDER).

\bibliography{LDPC,digwircom,Gmodels,gp,macler,PredisNLEq,svm,UMTS}
\bibliographystyle{IEEEtran}
\end{document}